# MP2-F12 Basis Set Convergence near the Complete Basis Set Limit: Are *h* Functions Sufficient?

Nisha Mehta and Jan M. L. Martin*

**ABSTRACT:** We have investigated the title question for the W4-08 thermochemical benchmark using *l*-saturated truncations of a large reference (REF) basis set, as well as for standard F12-optimized basis sets. With the REF basis set, the root-mean-square (RMS) contribution of *i* functions to the MP2-F12 total atomization energies (TAEs) is about 0.01 kcal/mol, the largest individual contributions being 0.04 kcal/mol for $P_2$ and $P_4$. However, even for these cases, basis set extrapolation from {*g,h*} basis sets adequately addresses the problem. Using basis sets insufficiently saturated in the *spdfgh* angular momenta may lead to exaggerated *i* function contributions. For extrapolation from *spdfg* and *spdfgh* basis sets, basis set convergence appears to be quite close to the theoretical asymptotic $\propto L^{-7}$ behavior. We hence conclude that *h* functions are sufficient even for highly demanding F12 applications. With one-parameter extrapolation, *spdf* and *spdfg* basis sets are adequate, aug-cc-pV{T,Q}Z-F12 yielding a RMSD = 0.03 kcal/mol. A limited exploration of CCSD(F12*) and CCSD-F12b suggests our conclusions are applicable to higher-level F12 methods as well.

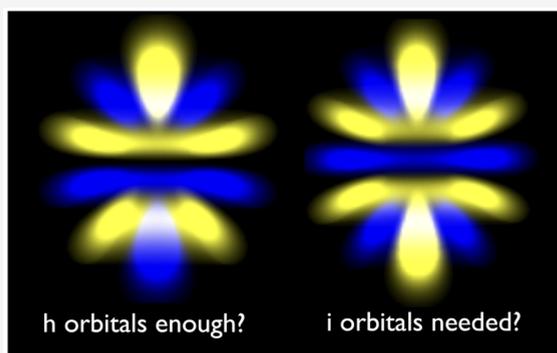

## ■ INTRODUCTION

Explicitly correlated quantum chemistry methods (see refs 1−3 for reviews) get their name from the inclusion of basis functions that involve explicit interelectronic distances (so-called "geminal" functions, as distinct from "orbital" functions that only involve a single electronic position).

The *de facto* standard at this point are the F12 geminals introduced by Ten-no:[4]

$$F(r_{12}) = \frac{1 - \exp(\gamma r_{12})}{\gamma} \quad (1)$$

which in the limit for small $r_{12}$ corresponds to adding $r_{12}$, and hence to satisfying the Kato cusp condition.[5] (For reasons of computational convenience, the Slater function is typically approximated by a linear combination of Gaussians, which actually is reminiscent of the Gaussian geminal approach of Persson and Taylor[6] a decade earlier.)

Exigencies for the underlying orbital basis set are quite different from those in a pure orbital calculation, as not so much effort needs to be invested in describing correlation near the interelectronic cusp. Hence basis sets specifically optimized for F12 calculations, such as the cc-pV*n*Z-F12 (*n* = D,T,Q,5) family by Peterson and co-workers,[7,8] or their anion-friendly aug-cc-pV*n*Z-F12 variants,[9] perform much better in an F12 setting than similar-sized basis sets for orbital calculations (such as the correlation consistent family of Dunning and co-workers). Indeed, both their contraction patterns and their exponents are markedly different[7] from those for the corresponding orbital-optimized basis sets. In fact, it has been shown[10] that non-F12 basis sets in thermochemical applications lead to erratic, nonmonotonous basis set convergence due to elevated basis set superposition error.

It is well-known that F12 calculations exhibit fairly rapid basis set convergence in terms of the maximum angular momentum *L* in the basis set: for two-electron model systems, Kutzelnigg and Morgan[11] showed $\propto L^{-7}$ for explicitly correlated calculations, compared to $L^{-3}$ for singlet-coupled pair correlation energies in pure orbital calculations.

Hence, few F12 practitioners go beyond *L* = 4 (i.e., *g* functions), that is, beyond cc-pVQZ-F12 or aug-cc-pVQZ-F12 basis sets. In accurate thermochemical applications, one may find (e.g., refs 10 and 12) cc-pV5Z-F12 or aug-cc-pwCV5Z[13−15] applied, both of which go up to *h* functions. At least one major electronic structure code often used for F12 calculations, MOLPRO,[16] has a hard limit of *i* functions overall, and hence (because of the need for one extra *L* step in the RI-MP2 auxiliary basis set) in practice the orbital basis set in F12 calculations tops out at *h* functions. Two other codes, Turbomole[17] and the most recent versions of ORCA, are capable of going up to at least $i$ functions in an F12 context.

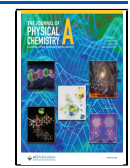

Table 1. RMSD (kcal/mol) for the TAEs of the W4-08 Dataset

| | truncated REF basis sets | | | | |
|---|---|---|---|---|---|
| | REF-d | REF-f | REF-g | REF-h | REF-i |
| | Ordinary MP2 | | | | |
| a | 13.136 | 4.965 | 2.205 | 1.203 | 0.789 |
| b | 13.137 | 4.981 | 2.221 | 1.219 | 0.791 |
| | MP2-F12 | | | | |
| a | 1.942 | 0.334 | 0.062 | 0.016 | 0.004 |
| b | 1.898 | 0.329 | 0.060 | 0.014 | 0.006 |
| | Extrapolations MP2-F12 | | | | |
| | REF-{d,f} | REF-{f,g} | REF-{g,h} | REF-{h,i} | |
| a | 0.133 | 0.027 | 0.006 | REF | |
| b | 0.127 | 0.022 | REF | 0.006 | |
| | F12 optimized correlation consistent | | | | |
| | VDZ-F12 | VTZ-F12 | AVTZ-F12 | VQZ-F12 | AVQZ-F12 | V5Z-F12 |
| a | 1.437 | 0.325 | 0.315 | 0.084 | 0.063 | 0.044 |
| b | 1.444 | 0.328 | 0.317 | 0.085 | 0.065 | 0.046 |
| | core−valence correlation consistent | | | | |
| | ACV6Z{f} | ACV6Z{g} | ACV6Z{h} | ACV6Z | awCV5Z | ditto, AV6Z(H) |
| a | 0.375 | 0.082 | 0.028 | 0.019 | 0.030 | 0.033 |
| b | 0.368 | 0.081 | 0.028 | 0.018 | 0.027 | 0.031 |
| | Valence aug-cc-pV(n + d)Z | | | | |
| | AV5Z | AV6Z(h) | AV6Z | | | |
| a | 0.055 | 0.041 | 0.020 | | | |
| b | 0.059 | 0.051 | 0.022 | | | |

[a]Relative to REF-{h,i} (largest systems omitted, 90 systems retained). [b]Relative to REF-{g,h} limit (all of W4-08, 96 systems).

$i$ functions are routinely employed in accurate *orbital-only* calculations, usually with basis set extrapolation as in the W4,[18−20] HEAT,[21,22] and FPD[23−25] thermochemical protocols. One study in our group[12] on the W4−17 thermochemical benchmark[26] went up to $k$ functions.

In the present note, we will investigate basis set convergence at the MP2-F12 level for the total atomization energies (TAEs) in the W4-08 subset[27] of W4-17. We will show that, while $i$ functions may still make some contributions to the atomization energies of some molecules (particularly, those featuring multiple bonds between second-row elements), this can be adequately addressed through $\propto L^{-7}$ basis set extrapolation, and *de facto* F12 basis set convergence is achieved with $h$ functions.

## ■ COMPUTATIONAL METHODS

All calculations were carried out using ORCA 5,[28−30] with density fitting for HF and MP2 disabled through the NoCoSX and NoRI keywords. This leaves only the CABS (complementary auxiliary basis set for F12) as a fitting basis set, thus eliminating the Coulomb-exchange and RI-MP2 fitting basis sets as possible "confounding factors". For technical reasons, UHF references were used for open-shell species.

The largest *spdfgh* basis set we considered was the reference basis set from Hill et al.[31] which was also used in refs 8 and 10, for calibrating the V5Z-F12 basis set. The *sp* part of this is the aug-cc-pV6Z (AV6Z for short) basis set from which two (first row) or one (second row) additional primitives were decontracted; the *dfgh* part is made up of large even-tempered sequences. This basis set we denote REF-h. For its CABS, we used the large uncontracted "reference-ri" basis set associated with it. REF-f and REF-g basis sets were generated by simple truncation of REF-h at $f$ and $g$ functions, respectively. (The linear dependency threshold for CABS was left at its default value of $10^{-8}$ throughout. The MP2-F12 ansatz used in ORCA corresponds to version D, which is a slight simplification[32] of ansatz C,[33,34] together with fixed geminal amplitudes[4,31] determined from the Kato cusp condition. This is basically equivalent to the default of "3C(Fix)" in MOLPRO.[16])

In addition, for a large subset of molecules, we considered an even larger REF-i basis set to which four $i$ functions have been added. Here, for the CABS basis set, we added a $k$ function with the same exponent to every $i$ function.

Aside from VTZ-F12, VQZ-F12, and V5Z-F12 basis sets, we considered aug-cc-pV(n+d)Z (n = T,Q,5,6; AVnZ+d for short),[35−38] as well as the core−valence correlation basis sets aug-cc-pwCVnZ (n = T,Q,5; awCVnZ for short),[15] and aug-cc-pCVnZ (n = Q,5,6; ACVnZ).[15] (We are only correlating valence electrons here, but it has been shown[8,10,12] that the additional radial flexibility of core−valence basis sets is beneficial for F12 calculations as well.) For VnZ-F12 (n = D,T,Q; VnZ-F12 for short) and AVnZ (n = T,Q,5), standard CABS "OptRI" basis sets[39,40] are available. For AV6Z+d we carried out two sets of calculations; in the first, we repurposed Hättig's unpublished DF-AV6Z basis set from the Turbomole library (downloaded from the Basis Set Exchange[41]) as the CABS; in the second, we employed the reference-RI from ref 31 instead. For the ACV6ZnoI basis set, which is a simple truncation of ACV6Z at $h$ functions, we employed reference-RI as the CABS.

As for the geminal exponent, for VTZ-F12 and VQZ-F12 we set $\gamma = 1.0$, and for V5Z-F12 $\gamma = 1.2$, as recommended in the literature for these respective basis sets.[7,8] For the REF-n and ACVnZ, a fixed $\gamma = 1.4$ was used throughout as per common practice for large orbital basis sets. (An elevated gamma restricts the geminal to the closer-in part of the cusp.)

Table 2. TAE (kcal/mol) as a Function of the Basis Set for Some Representative Molecules

| | MP2 | MP2-F12 truncated REF basis sets | | | | | differences | | | extrapolations | | | |
|---|---|---|---|---|---|---|---|---|---|---|---|---|---|
| basis | AV{5,6}Z + d | REF-d | REF-f | REF-g | REF-h | REF-i | $\Delta h^a$ | $\Delta i^b$ | {h,i} − {g,h} | {d,f} | {f,g} | {g,h} | {h,i} |
| BF$_3$ | 496.38 | 496.45 | 496.56 | 496.53 | 496.52 | 496.52 | −0.01 | 0.00 | 0.00 | 496.58 | 496.53 | 496.52 | 496.52 |
| AlCl$_3$ | 330.07 | 326.78 | 329.82 | 330.35 | 330.40 | 330.40 | 0.05 | 0.00 | −0.01$_7$ | 330.39 | 330.46 | 330.41$_4$ | 330.39$_7$ |
| S$_4$ | 260.73 | 252.68 | 259.49 | 260.53 | 260.66 | 260.67 | 0.13 | 0.01 | −0.02 | 260.77 | 260.74 | 260.70 | 260.68 |
| Si$_2$H$_6$ | 526.53 | 528.99 | 526.94 | 526.49 | 526.36 | 526.31 | −0.13 | −0.04 | −0.02 | 526.56 | 526.39 | 526.32 | 526.30 |
| P$_4$ | 308.18 | 296.18 | 306.29 | 307.91 | 308.14 | 308.18 | 0.23 | 0.04 | −0.01 | 308.20 | 308.23 | 308.21 | 308.20 |
| P$_2$ | 120.44 | 116.67 | 119.42 | 120.26 | 120.44 | 120.48 | 0.18 | 0.04 | 0.00 | 119.94 | 120.43 | 120.50 | 120.50 |
| Cl$_2$ | 64.49 | 61.80 | 64.24 | 64.52 | 64.55 | 64.56 | 0.03 | 0.01 | 0.00 | 64.70 | 64.57 | 64.56 | 64.56 |
| SO$_3$ | 382.96 | 383.30 | 382.82 | 382.61 | 382.53 | 382.51 | −0.07 | −0.02 | −0.01 | 382.73 | 382.57 | 382.51 | 382.50 |
| C$_2$H$_2$ | 414.26 | 413.81 | 414.16 | 414.25 | 414.26 | 414.26 | 0.01 | 0.00 | 0.00 | 414.22 | 414.26 | 414.26 | 414.26 |
| C$_2$H$_4$ | 566.53 | 567.32 | 566.64 | 566.51 | 566.48 | 566.47 | −0.03 | −0.01 | 0.00 | 566.52 | 566.48 | 566.47 | 566.47 |
| C$_2$H$_6$ | 712.78 | 713.87 | 712.91 | 712.74 | 712.70 | N/A | −0.04 | N/A | N/A | 712.73 | 712.70 | 712.69 | N/A |
| CO$_2$ | 415.47 | 415.17 | 415.47 | 415.56 | 415.57 | 415.57 | 0.01 | 0.00 | 0.00 | 415.53 | 415.57 | 415.57 | 415.57 |
| N$_2$O | 296.09 | 295.34 | 295.99 | 296.17 | 296.21 | 296.22 | 0.04 | 0.02 | 0.01 | 296.11 | 296.20 | 296.22 | 296.23 |
| SiH$_4$ | 318.13 | 319.74 | 318.38 | 318.08 | 318.00 | 317.98 | −0.08 | −0.02 | −0.01 | 318.13 | 318.02 | 317.98 | 317.97 |
| PH$_3$ | 234.51 | 236.24 | 234.75 | 234.46 | 234.40 | 234.38 | −0.07 | −0.02 | −0.01 | 234.47 | 234.40 | 234.38 | 234.37 |
| CH$_4$ | 417.87 | 418.47 | 417.93 | 417.84 | 417.82 | 417.81 | −0.02 | 0.00 | 0.00 | 417.83 | 417.82 | 417.81 | 417.81 |
| H$_2$O | 237.53 | 237.98 | 237.63 | 237.56 | 237.56 | 237.56 | −0.01 | 0.00 | 0.00 | 237.56 | 237.56 | 237.56 | 237.56 |

$^a\Delta h$ = TAE[REF-h] − TAE[REF-g]. $^b\Delta i$ = TAE[REF-i] − TAE[REF-h].

Throughout this paper, we will refer to two-point basis set extrapolation using the braces notation, for example, V{T,Q}Z-F12 refers to extrapolation from VTZ-F12 and VQZ-F12 basis sets. See ref 42 for a discussion on how all two-point extrapolation schemes (e.g., refs 43–46) are interrelated.

## ■ RESULTS AND DISCUSSION

**Convergence for l-Saturated Basis Sets through l = 6, REF-i.** Performance statistics with truncations of the large REF-i basis set at different angular momenta can be found in Table 1. We were able to obtain TAEs for all 96 species in W4-08 through REF-h, plus REF-i for all species except six: B$_2$H$_6$, C$_2$H$_6$, CH$_3$NH$_2$, CH$_2$NH$_2$, CH$_3$NH, and NCCN (dicyanogen).

As mentioned in the introduction, according to Kutzelnigg and Morgan,[11] the l-saturated basis set convergence behavior in an R12 calculation, for a closed-shell pair energy, should asymptotically be ∝ $L^{-7}$. Is this indeed the case here? One simple test would be to consider the RMS difference between MP2-F12 atomization energies obtained by $A + B \cdot L^{-7}$ extrapolation from the REF-{g,h} basis set pair with those obtained in the same manner from the REF-{h,i} pair. As can be seen in Table 1, the RMS difference between these two columns is just 0.006 kcal/mol. We obtain the same value to three decimal places if we minimize RMSD with respect to the REF-{g,h} extrapolation exponent, for which we find 6.514 as the optimum value. Both observations indicate that for these large l-saturated basis sets we are essentially in the ∝ $L^{-7}$ regime (note that even for REF-{f,g} we already obtain 6.232 as the optimum extrapolation exponent). Incidentally, they also suggest that the MP2-F12/REF-{h,i} values are as close as we can reasonably hope to get to a CBS limit reference. The largest individual differences are for cyclic S$_4$, 0.020 kcal/mol, and Si$_2$H$_6$, −0.020 kcal/mol, followed by 0.017 kcal/mol for AlCl$_3$, 0.0152 kcal/mol for N$_2$, and 0.012 kcal/mol for CS$_2$, SiH$_4$, and N$_2$O. Convergence for some representative molecules is summarized in Table 2.

Using MP2-F12/REF-{l−1,l} extrapolation, and minimizing RMSD with respect to the extrapolation exponents, we obtain REF-{d,f} 0.13, REF-{f,g} 0.027, and as already mentioned REF-{g,h} 0.006 kcal/mol; with the respective exponents 4.543, 6.232, and 6.514.

Let us now consider the TAE increments $\Delta h$ = TAE[MP2-F12/REF-h] − TAE[MP2-F12/REF-g] and $\Delta i$ = TAE[MP2-F12/REF-i] − TAE[MP2-F12/REF-h]. The RMS $\Delta h$ = 0.046 kcal/mol, but individual values can reach as high as 0.23 kcal/mol for P$_4$, 0.18 kcal/mol for P$_2$, and 0.13 kcal/mol for S$_4$. $\Delta i$ is much smaller, 0.011 kcal/mol RMS, but for P$_4$ and P$_2$ it reaches 0.042 and 0.039 kcal/mol, respectively. The latter are definitely amounts of interest for high-accuracy electronic structure calculations (as the MP2-F12 basis set incompleteness error is at least a semiquantitative indication of what the corresponding CCSD(T)-F12 error would be). For some perspective, however, the corresponding numbers for *orbital-only* MP2 are RMS($\Delta h$) = 1.01 and RMS($\Delta i$) = 0.43 kcal/mol, with the largest individual values (again for P$_4$) being 3.30 and 1.57 kcal/mol, respectively, followed by S$_4$, 2.63 and 1.13 kcal/mol. It is thus clear that MP2-F12 $\Delta i$ values are 1−1.5 orders of magnitude smaller, and that basis sets convergence has essentially been achieved by the time one gets to REF-h. For the anomalous cases of P$_2$ and P$_4$, the unusually slow basis set convergence has been documented in detail by Persson and Taylor[47] (see also Karton and Martin[48]). And even for those, $L^{-7}$ basis set extrapolation can clearly "fill in the cracks": the REF-{h,i} − REF-{g,h} difference for P$_4$ is just 0.011 kcal/mol, while that for P$_2$ vanishes entirely. We conclude that, at least for REF-l l saturated basis sets, there appears to be no compelling need to go beyond h functions in F12 calculations. This is good news, of course, for users of F12 correlation codes in program suites such as MOLPRO that do not permit going beyond i functions in the auxiliary basis sets (i.e., beyond h functions in the orbital basis set).

The slower basis set convergence for multiple-second-row species like P$_2$, P$_4$, S$_3$, and AlCl$_3$ compared to their isovalent first-row counterparts N$_2$, N$_4$, O$_3$, and BF$_3$, respectively, can be rationalized to some degree in terms of the lower-lying 3d (and, to a lesser degree, 4f orbitals). This is probably best illustrated by considering the d, f, and g populations in a natural population analysis (NPA),[50] presented in Table 3 for selected molecules at the HF/AVQZ+d and CCSD(T)/AVQZ+d levels. One sees a high d population already at the HF level for cases like $SO_3$, which is a paradigmatic example of "inner polarization" in which the oxygen lone pairs back-donate into the empty 3d of sulfur (and similarly for other second row elements in high oxidation states; see ref 51 and references therein), which thereby becomes an 'honorary valence orbital of the second kind'.[52] It should be noted that this is primarily an SCF-level effect, and hence the increase in 3d population upon introducing correlation is quite modest in comparison. For cases like $P_2$ and $P_4$, however, this effect is clearly not operative—one does see a significant d population even at the HF level, but it is much enhanced by correlation, especially when compared to the isovalent first-row species $N_2$ and $N_4$. Moreover, this is not just limited to the d orbitals: natural orbital occupations for f and g shells clearly decay more slowly for the second-row species than for their first-row cognates. Similar observations can be made for $Cl_2$ vs $F_2$, $S_3$ vs $O_3$, and the like.

Table 3. NBO Angular Momentum Populations at the HF/cc-pV(Q+d)Z and CCSD(T)/cc-pV(Q+d)Z Levels for Selected Molecules[a]

| | | HF/aug-cc-pV(Q+d)Z | | | CCSD(T)/aug-cc-pV(Q+d)Z | | |
|---|---|---|---|---|---|---|---|
| | | d | f | g | d | f | g |
| $N_2$ | N | 0.02493 | 0.00186 | 0.00047 | 0.04513 | 0.00509 | 0.00110 |
| $P_2$ | P | 0.05648 | 0.00407 | 0.00127 | 0.11985 | 0.01390 | 0.00295 |
| $N_4$ | N | 0.02957 | 0.00297 | 0.00177 | 0.05414 | 0.00719 | 0.00322 |
| $P_4$ | P | 0.07726 | 0.00759 | 0.00231 | 0.14830 | 0.01982 | 0.00591 |
| $F_2$ | F | 0.00417 | 0.00036 | 0.00008 | 0.03159 | 0.00442 | 0.00079 |
| $Cl_2$ | Cl | 0.02517 | 0.00232 | 0.00029 | 0.13406 | 0.01664 | 0.00315 |
| ClF | Cl | 0.02472 | 0.00114 | 0.00009 | 0.12730 | 0.01451 | 0.00250 |
| | F | 0.01462 | 0.00159 | 0.00014 | 0.04186 | 0.00607 | 0.00105 |
| $BF_3$ | B | 0.01056 | 0.00148 | 0.00133 | 0.02739 | 0.00487 | 0.00227 |
| | F | 0.01533 | 0.00089 | 0.00014 | 0.04272 | 0.00457 | 0.00090 |
| $AlF_3$ | Al | 0.03366 | 0.00374 | 0.00131 | 0.05659 | 0.00763 | 0.0020 |
| | F | 0.01551 | 0.00055 | 0.00005 | 0.04518 | 0.00445 | 0.00092 |
| $AlCl_3$ | Al | 0.04607 | 0.00326 | 0.00126 | 0.07284 | 0.01325 | 0.00539 |
| | Cl | 0.02818 | 0.00199 | 0.00033 | 0.14337 | 0.01487 | 0.00303 |
| $O_3$ | $O_{center}$ | 0.02316 | 0.00262 | 0.00139 | 0.05121 | 0.00824 | 0.00280 |
| | $O_{arm}$ | 0.00932 | 0.00077 | 0.00016 | 0.03595 | 0.00464 | 0.00092 |
| $SO_2$ | S | 0.13859 | 0.00462 | 0.00158 | 0.20438 | 0.01357 | 0.00340 |
| | O | 0.05272 | 0.00284 | 0.00037 | 0.07735 | 0.00736 | 0.00135 |
| $S_3$ | $S_{center}$ | 0.13475 | 0.00947 | 0.00350 | 0.21280 | 0.02570 | 0.00712 |
| | $S_{arm}$ | 0.03427 | 0.00290 | 0.00042 | 0.12586 | 0.01544 | 0.00274 |
| $CF_2$ | C | 0.01101 | 0.00076 | 0.00035 | 0.03559 | 0.00379 | 0.00106 |
| | F | 0.01858 | 0.00144 | 0.0002 | 0.04511 | 0.00540 | 0.00106 |
| $CCl_2$ | C | 0.02856 | 0.00389 | 0.00247 | 0.06138 | 0.01049 | 0.00395 |
| | Cl | 0.02118 | 0.00187 | 0.00032 | 0.12422 | 0.01451 | 0.00294 |
| $SO_3$ | S | 0.20747 | 0.00753 | 0.00531 | 0.26897 | 0.01818 | 0.00714 |
| | O | 0.04415 | 0.00249 | 0.00039 | 0.06929 | 0.00673 | 0.00128 |
| $S_4$ | $S_{bridge}$ | 0.08992 | 0.00638 | 0.00186 | 0.17087 | 0.02129 | 0.00515 |
| | $S_{apex}$ | 0.03694 | 0.00354 | 0.00058 | 0.13287 | 0.01680 | 0.00311 |
| $F^-$ | | 0.0 | 0.0 | 0.0 | 0.03380 | 0.00428 | 0.00086 |
| $Cl^-$ | | 0.0 | 0.0 | 0.0 | 0.12942 | 0.01480 | 0.00335 |
| HF | F | 0.00987 | 0.00049 | 0.00006 | 0.03680 | 0.00404 | 0.00078 |
| HCl | Cl | 0.01563 | 0.00089 | 0.00006 | 0.12527 | 0.01291 | 0.00252 |

[a]Populations refer to unique atoms (e.g., to one of the four equivalent N atoms in tetrahedral $N_4$). CCSD(T)/cc-pVQZ geometry for $N_4$, $r_{NN}$ = 1.4551 Å, was taken from ref 49.

**cc-pVnZ-F12 and Related Basis Sets.** How does the cc-pVnZ-F12 series perform compared to the REF-{h,i} basis set limit estimate? RMSDs are 1.44, 0.33, 0.084, and 0.044 for n = D,T,Q,5, respectively. The latter is close to the "no extrapolation required" goal, but there is still some room for improvement in high-accuracy thermochemistry applications, where one would like a $3\sigma$ = 0.1 kcal/mol error bar. AVTZ-F12 performs only marginally better (0.31 kcal/mol) than the underlying VTZ-F12, while AVQZ-F12, at 0.063 kcal/mol, does represent a modest gain over VQZ-F12. It was previously shown[8,12] that the aug-cc-pwCV5Z core–valence basis set (awCV5Z for short) performs remarkably well for F12 calculations, despite not being developed for this purpose at all: the RMSD = 0.030 kcal/mol we find here is consistent with that observation. (Expanding the hydrogen basis set from AV5Z to AV6Z yields the same performance to within statistical noise, RMSD = 0.033 kcal/mol.) The underlying valence-correlation basis set, aug-cc-pV5Z or AV5Z, has a higher RMSD = 0.055 kcal/mol; adding some radial flexibility by instead using AV6Z(h), that is, aug-cc-pV6Z with the highest angular momentum i removed, reduces RMSD slightly to 0.041 kcal/mol. This however is cut in half when the i functions are restored, RMSD = 0.022 kcal/mol for AV6Z. By comparing the two columns of numbers, we find several molecules where i function contributions deceptively seem to reach 0.1 kcal/mol, such as $P_4$, $S_4$, and $AlCl_3$, plus another 0.07 kcal/mol for $S_3$. However, we have already established with the REF-h and REF-i basis set that the true $i$ contribution is much smaller; it is a well-known basis set convergence phenomenon (see, e.g., the older review of basis sets by Davidson and Feller[53]) that insufficiently saturating the basis in lower angular momenta can exaggerate the impact of the highest angular momentum (through basis set superposition error). If we instead compare the core−valence versions of these basis sets, the $i$-increments are substantially reduced, with $P_4$ and $P_2$ now remaining as the principal outliers. Finally, if one wants a single "prepackaged" basis set without extrapolation, ACV6Z has the smallest RMSD (0.019 kcal/mol) of them all.

**A Comment about Basis Set Extrapolation.** We have considered three sets of extrapolation exponents:

1. From the literature, as optimized for the very small training set of Hill et al.[31] (that paper itself for V{D,T}Z-F12 and V{T,Q}Z-F12), ref [54] for V{Q,5}Z-F12, and ref [55] for AV{T,Q}Z-F12.
2. By minimization of RMSD from REF-{h,i} total energies for such W4-08 species for which we have REF-i energies.
3. By minimization of RMSD from REF-{h,i} TAEs for the same species.

The three sets of extrapolation exponents are compared in Table 4 below, as are the RMSDs for total atomization energies with them. The TAE-optimized exponent V{Q,5}Z-F12 appears anomalously small, but this is an artifact of the very flat surface there, which precludes a "clean" optimization. $E_{total}$ gives a somewhat better defined minimum, though even there the surface is quite flat. In terms of RMSD, for V{Q,5}Z-F12 all three possible exponents yield the same error. For AV{T,Q}Z-F12, both optimized values from the present work yield superior RMSDs to the value from ref [55]; for VTZ-F12 the same holds, although the difference here is quite modest. Finally, for the V{D,T}Z-F12 pair, the two optimized values from the present work are clearly superior to the one from Hill et al.[31] For reasons of numerical stability (notably because it eliminates the anomaly for V{Q,5}Z-F12), we favor the set optimized from total energies. The difference in RMSD is negligible; however, this is good news, since ideally one wants to be in a scenario where basis set extrapolation is only a minor component of the final result and is relatively insensitive to details of the extrapolation procedure. Note that, while all basis set pairs considered here are clearly some distance away from the asymptotic $L^{-7}$ regime, the exponents do increase in that direction as $L$ gets larger.

**Table 4. Two-Point Extrapolation Exponents and RMSD from REF-{h,i} (kcal/mol) for Different Basis Set Pairs**

|  | V{D,T}Z-F12 | V{T,Q}Z-F12 | AV{T,Q}Z-F12 | V{Q,5}Z-F12 |
|---|---|---|---|---|
| *Extrapolation Exponents* | | | | |
| lit.[31,54,55] | 3.0878 | 4.3548 | 4.6324 | 5.0723 |
| W4-08 $E_{total}$ | 3.45 | 5.03 | 5.64 | 5.27 |
| W4-08 TAE | 3.80 | 5.11 | 5.97 | 4.36 |
| *RMSD with These Exponents (kcal/mol)* | | | | |
| lit. | 0.178 | 0.046 | 0.046 | 0.034 |
| W4-08 $E_{total}$ | 0.123 | 0.039 | 0.030 | 0.034 |
| W4-08 TAE | 0.107 | 0.038 | 0.029 | 0.033 |

|  | AV{D,T}Z+d | AV{T,Q}Z+d | AV{Q,5}Z+d | AV{5,6}Z+d |
|---|---|---|---|---|
| *regular MP2 for comparison* | | | | |
| *Extrapolation Exponents* | | | | |
| lit.[a] | 2.136 | 2.531 | 2.740 | 2.835 |
| W4-08 TAE | 2.571 | 2.963 | 2.908 | 2.910 |
| *RMSD with These Exponents (kcal/mol)* | | | | |
| lit.[a] | 3.386 | 1.062 | 0.272 | 0.118 |
| W4-08 TAE | 1.800 | 0.458 | 0.216 | 0.111 |

[a]Extrapolation exponents from Table 8 of ref [31].

For some perspective, we add the RMSDs for regular MP2 with aug-cc-pV($L$−1+d)Z and aug-cc-pV($L$+d)Z basis sets. It is sobering to see that even AV{5,6}Z+d only reaches the accuracy level of V{D,T}Z-F12, and that V{T,Q}Z-F12 already markedly exceeds it.

**What Do These Results Imply for CCSD-F12 and CCSD(T)-F12?** At the request of a reviewer, we will address what we may infer for the basis set convergence of higher-level methods like CCSD(T)(F12*) or CCSD(T)-F12b.

First of all, the treatment of parenthetical triples (T) does not benefit in any way from the F12 treatment,[56] so the convergence of that contribution is effectively the same as in an orbital calculation. The latter has been addressed in great detail in a recent book chapter by one of us:[57] suffice to say here that basis set convergence of (T) with angular momentum is actually fairly rapid, and that radial flexibility of the basis set is actually more important than angular flexibility.

This leaves us then with CCSD-F12, or rather with the various practical approximations to it such as CCSD-F12b,[58] $CCSD_{\overline{F12}}$,[59,60] and CCSD(F12*).[61] Of these, CCSD(F12*) has been shown[62] to be the most rigorous approximation that still is computationally feasible, while CCSD-F12b is widely used owing to its implementation for both closed-shell and open-shell cases in the MOLPRO[16] program system.

The basis set convergence of differences between different CCSD-F12 approximations has been studied in great detail in ref [12]. In a nutshell: if correlation is predominantly dynamic then these differences decay quickly with increasing basis sets, but even moderate amounts of static correlation can cause nontrivial differences (as large as 0.3 kcal/mol) to persist even for *spdfgh* basis sets.

As additional complications, all of the available (to us) implementations of approximate CCSD-F12 methods require DFMP2 and JKFit auxiliary basis sets (further complicating comparisons), and the only CCSD(F12*) implementation at our disposal with which we were able to get enough calculations in REF-h basis sets to converge, i.e., that in MOLPRO, is limited to closed shell. Hence we resorted to calculating a number of closed-shell reaction energies instead. The relevant basis set increments are reported in Table 5. Auxiliary basis sets were taken from the Supporting Information of ref [31].

The basis set increments given there for $g$ and $h$ layers are obtained directly. For the $i$ layer, the ORCA MP2-F12 values are calculated directly while the MOLPRO values are obtained by $L^{-\alpha}$ extrapolation. (It can be easily shown that an estimate for the next basis set increment after $E(L) - E(L - 1)$ is given by the following formula:)

$$E(L + 1) - E(L) \approx [E(L) - E(L - 1)] \frac{1 - \left(\frac{L}{L+1}\right)^\alpha}{\left(\frac{L}{L-1}\right)^\alpha - 1} \quad (2)$$

Table 5. Comparison of REF-*n* Basis Set Increments (kcal/mol) for a Number of Closed-Shell Reactions at the MP2-F12, CCSD-F12b, and CCSD(F12*) Levels[a]

|  |  | ORCA | MOLPRO (includes MP2Fit and JKFit) | | | |
|---|---|---|---|---|---|---|
|  |  | MP2-F12 | MP2-F12 | CCSD(F12*) | CCSD-F12b | (F12*)-F12b |
| $P_4 \rightarrow 2P_2$ | $\Delta g$ | 0.054 | 0.066 | −0.014 | −0.125 | 0.111 |
|  | $\Delta h$ | 0.132 | 0.108 | 0.006 | −0.042 | 0.048 |
|  | $\Delta i$ | 0.036 | 0.026(5) | 0.001(0) | 0.010(2) | −0.009(2) |
| $2HF \rightarrow F_2 + H_2$ | $\Delta g$ | 0.078 | 0.076 | 0.089 | −0.014 | 0.103 |
|  | $\Delta h$ | 0.011 | 0.008 | 0.021 | −0.012 | 0.033 |
|  | $\Delta i$ | 0.002 | 0.002(0) | 0.005(1) | 0.003(1) | 0.002(1) |
| $CO + H_2O \rightarrow CO_2 + H_2$ | $\Delta g$ | 0.127 | 0.126 | 0.180 | 0.317 | −0.137 |
|  | $\Delta h$ | 0.017 | 0.014 | 0.011 | 0.036 | −0.025 |
|  | $\Delta i$ | 0.003 | 0.003(1) | 0.003(1) | 0.009(2) | −0.006(2) |
| $N_2 + H_2O \rightarrow N_2O + H_2$ | $\Delta g$ | 0.139 | 0.14 | 0.173 | 0.245 | −0.072 |
|  | $\Delta h$ | 0.022 | 0.015 | 0.014 | 0.022 | −0.008 |
|  | $\Delta i$ | 0.002 | 0.004(1) | 0.003(1) | 0.005(1) | −0.002(0) |
| $3H_2S \rightarrow S_3 + 3H_2$ | $\Delta g$ | 2.026 | 2.038 | 1.716 | 2.013 | −0.297 |
|  | $\Delta h$ | 0.380 | 0.373 | 0.15 | 0.213 | −0.063 |
|  | $\Delta i$ | 0.082 | 0.090(19) | 0.036(7) | 0.051(11) | −0.015(3) |
| $H_2S + 2H_2O \rightarrow SO_2 + 3H_2$ | $\Delta g$ | 0.354 | 0.359 | 0.433 | 0.626 | −0.193 |
|  | $\Delta h$ | 0.045 | 0.038 | 0.052 | 0.099 | −0.047 |
|  | $\Delta i$ | 0.006 | 0.009(2) | 0.012(3) | 0.024(5) | −0.012(3) |
| $2PH_3 \rightarrow P_2 + 3H_2$ | $\Delta g$ | 1.420 | 1.424 | 0.923 | 1.056 | −0.133 |
|  | $\Delta h$ | 0.315 | 0.31 | 0.085 | 0.109 | −0.024 |
|  | $\Delta i$ | 0.075 | 0.075(16) | 0.021(4) | 0.026(5) | −0.005(1) |
| $4PH_3 \rightarrow P_4 + 6H_2$ | $\Delta g$ | 2.786 | 2.782 | 1.859 | 2.236 | −0.377 |
|  | $\Delta h$ | 0.499 | 0.511 | 0.165 | 0.259 | −0.094 |
|  | $\Delta i$ | 0.113 | 0.123(26) | 0.040(8) | 0.063(13) | −0.023(5) |
| $3H_2O \rightarrow O_3 + 3H_2$ | $\Delta g$ | 0.229 | 0.233 | 0.243 | 0.137 | 0.106 |
|  | $\Delta h$ | 0.039 | 0.032 | 0.042 | −0.013 | 0.055 |
|  | $\Delta i$ | 0.008 | 0.008(2) | 0.010(2) | 0.003(1) | 0.007(2) |

[a]Values that do not include an uncertainty interval in parentheses were calculated directly. Values that do include such an uncertainty interval were estimated as the average between $L^{-7}$ (ideal case) and $L^{-5}$ (pessimistic scenario) extrapolation, with one-half the distance between the two values taken as the uncertainty interval.

For the sake of the estimate, we considered $\alpha = 7$ (the asymptotic convergence rate) as a best-case scenario, and $\alpha = 5$ as a worst-case scenario (for VQZ-F12 and V5Z-F12, we have previously found[54] the intermediate value $\alpha \approx 6$). The average of both extrapolated values, plus or minus half the difference between them, is taken as our approximate estimate.

As can be seen in Table 5, basis set convergence of especially CCSD(F12*) actually seems, depending on the reaction, comparable to or faster than that of MP2-F12. And for those reactions where $\Delta i$ is still somewhat significant at the MP2-F12 level, the corresponding estimated CCSD-F12b and especially CCSD(F12*) increments are actually up to 3 times smaller.

We hence conclude that our conclusions about the lack of necessity of *i* functions in MP2-F12 calculations also apply to CCSD-F12b, CCSD(F12*), and other approximations to CCSD-F12.

As an aside, we note that the CCSD(F12*) − CCSD-F12b differences, while nontrivial, are conspicuously smaller than what was observed for cc-pVQZ-F12, cc-pV5Z-F12, and aug-cc-pwCV5Z in ref 12. Clearly, here too, the greater radial flexibility of the REF-*n* basis sets puts them at an advantage, even as they are unwieldy for practical production calculations.

## ■ CONCLUSIONS

We may conclude that for basis sets that are adequately saturated in the angular momenta represented, the contribution of *i* functions to total atomization energies is minimal (about 0.01 kcal/mol RMS): the largest individual contributions we have found here only reach 0.04 kcal/mol (for $P_4$ and $P_2$), and even that is adequately captured by $L^{-7}$ extrapolation from REF-{g,h}, which is closer than 0.01 kcal/mol RMS to the CBS limit. (In this $L$ region, basis set convergence behavior for MP2-F12 is found to be quite close to the asymptotic[11] $L^{-7}$.) Comparison of AV6Z with and without *i* functions indicates more significant *i* contributions, owing to insufficient radial saturation. The *h* contribution is more significant, reaching 0.23 kcal/mol for $P_4$. REF-{g,h} extrapolation is within less than 0.01 kcal/mol RMS of REF-{h,i}, the largest single difference being 0.02 kcal/mol for $S_4$. This means that the REF-{g,h} $L^{-7}$ extrapolation is an acceptable proxy for the F12 basis set limit, and that *h* functions are sufficient even for highly demanding F12 applications.

If one-parameter extrapolation with an adjustable exponent is permissible, *spdf* and *spdfg* basis sets are adequate, with aug-cc-pV{T,Q}Z-F12 yielding the RMSD = 0.03 kcal/mol.

Finally, the somewhat slower basis set convergence in the second row—especially when there are multiple adjacent such elements—can be rationalized to some degree in terms of lower-lying 3*d* and 4*f* orbitals.

A limited investigation for closed-shell reactions indicates that basis set convergence of CCSD(F12*) and CCSD-F12b is comparable to or faster than that of MP2-F12, and that hence our conclusions about the relative insignificance of *i*-function contributions apply to F12 methods more broadly.

## AUTHOR INFORMATION

**Corresponding Author**

Jan M. L. Martin − *Department of Molecular Chemistry and Materials Science, Weizmann Institute of Science, Reḥovot 7610001, Israel;* orcid.org/0000-0002-0005-5074; Phone: +972-8-9342533; Email: gershom@weizmann.ac.il; Fax: +972-8-9343029

**Author**

Nisha Mehta − *Department of Molecular Chemistry and Materials Science, Weizmann Institute of Science, Reḥovot 7610001, Israel;* orcid.org/0000-0001-7222-4108

Complete contact information is available at: https://pubs.acs.org/10.1021/acs.jpca.2c02494

**Notes**

The authors declare no competing financial interest.

## ACKNOWLEDGMENTS

Work on this paper was supported in part by the Israel Science Foundation (Grant 1969/20) and by the Minerva Foundation (Grant 2020/05). We thank Prof. Amir Karton for helpful discussions. This paper is dedicated to Prof. Peter R. Taylor in honor of his 70th birthday, and of his landmark contributions (e.g., refs 6 and 63−65) to the theory and practice of explicitly correlated quantum chemistry.